\documentclass[a4paper]{jpconf}
\usepackage{latexsym}
\usepackage{graphicx}
\begin{document}
\title{Lithium Diffusion \& Magnetism in\\Battery Cathode Material Li$_{\rm x}$Ni$_{1/3}$Co$_{1/3}$Mn$_{1/3}$O$_{2}$}

\author{M. M\aa nsson$^{1,2}$, H. Nozaki$^3$, J.~M.~Wikberg$^4$, K. Pr\v{s}a$^{1,5}$, Y. Sassa$^{5}$, \\M.~Dahbi$^6$, K. Kamazawa$^7$, K. Sedlak$^8$, I. Watanabe$^9$, J. Sugiyama$^3$ }

\address{$^1$ Laboratory for Quantum Magnetism, $\acute{\rm E}$cole Polytechnique F$\acute{\rm e}$d$\acute{\rm e}$rale de Lausanne (EPFL),
CH-1015 Lausanne, Switzerland}
\address{$^2$ Laboratory for Neutron Scattering \& Imaging, Paul Scherrer Institute, CH-5232 Villigen PSI, Switzerland}
\address{$^3$ Toyota Central Research \& Development Laboratories, Inc., 41-1 Yokomichi, Nagakute, Aichi 480-1192, Japan}
\address{$^4$ Department of Physics and Astronomy, Molecular and Condensed Matter Physics, Uppsala University, Box 516, S-75120 Uppsala, Sweden}
\address{$^5$ Laboratory for Solid state physics, ETH Z\"{u}rich, CH-8093 Z\"{u}rich, Switzerland}
\address{$^6$ Department of Chemistry, {\AA}ngstr\"{o}m Laboratory, Uppsala University, Box 538, S-75121 Uppsala, Sweden}
\address{$^7$ Comprehensive Research Organization for Science and Society (CROSS), Tokai, Ibaragi 319-1106, Japan}
\address{$^8$ Laboratory for Muon Spin Spectroscopy, Paul Scherrer Institute, CH-5232 Villigen PSI, Switzerland}
\address{$^9$ Advanced Meson Science Laboratory, RIKEN, 2-1 Hirosawa, Wako, Saitama 351-0198, Japan}

\ead{martin.mansson@epfl.ch}

\begin{abstract}
We have studied low-temperature magnetic properties as well as high-temperature lithium ion diffusion in the battery cathode materials Li$_{\rm x}$Ni$_{1/3}$Co$_{1/3}$Mn$_{1/3}$O$_{2}$ by the use of muon spin rotation/relaxation. Our data reveal that the samples enter into a 2D spin-glass state below $T_{\rm SG}\approx12$~K. We further show that lithium diffusion channels become active for $T\geq T_{\rm diff}\approx125$~K where the Li-ion hopping-rate [$\nu(T)$] starts to increase exponentially. Further, $\nu(T)$ is found to fit very well to an Arrhenius type equation and the activation energy for the diffusion process is extracted as $E_a\approx100$~meV.
\end{abstract}

\section{Introduction}
One of the main obstacles for a general breakthrough of electric automobiles is the development of a high-capacity, cheap and safe rechargeable battery. The most widely used cathode material is by far LiCoO$_{2}$ \cite{Mizushima,Li_review}, however, cobalt is very expensive and there is a strong driving force to find new cheaper and more environmental friendly cathode materials. One of the big problems for battery application is the structural distortions that occur \cite{VanderVen,Seguin} when large amounts of Li is deintercalated during charge / discharge cycles of the battery [see Fig.~1]. Some efforts have been made to minimize these effects $e.g.$ by combining Co and Ni into a LiNi$_{1-x}$Co$_{x}$O$_{2}$ solid solution. However, this compound suffers from strong intermixing of Ni into the Li layers for the highly delithiated state \cite{Chebiam}. Recently, the mixed transition metal oxides (MTMO's) of the form
Li$_{\rm x}$Ni$_{1-y-z}$Co$_{y}$Mn$_{z}$O$_{2}$ have become the center of attention \cite{Yoshio,Choi}. In particular has the Li$_{\rm x}$Ni$_{1/3}$Co$_{1/3}$Mn$_{1/3}$O$_{2}$ compound \cite{Ohzuku} been put forward as one of the most promising battery electrode materials for the future. The structure of these compounds is the same as for the fundamental LiCoO$_{2}$ i.e. a rhomohedral lattice (space group $R\overline{3}m$) where Ni/Co/MnO$_{2}$ planes are stacked between nonmagnetic Li layers along the $c$-axis [see Fig.~1]. In similarity with LiNiO$_{2}$ some intermixing of Ni in the Li layers can be expected. However, in this particular case, this actually helps to stabilize the structure and improve capacity retention \cite{Dahbia}.

In addition to their extensive use in energy related applications, these materials also display very interesting and complex magnetic properties at lower temperatures. This is due to the 2D triangular antiferromagnetic (AF) lattice \cite{Anderson} (see also Fig.~1) created by the transition metal ions. For the Li$_{\rm x}$Ni$_{1/3}$Co$_{1/3}$Mn$_{1/3}$O$_{2}$ compounds the situation is even more complex since a combination of three different transition metals is present within the geometrically frustrated magnetic planes \cite{Wikberg_PRB,Wikberg_JAP,Wikberg_muSR}. In this paper we have investigated both the low-temperature magnetic properties as well as the high-temperature ion-diffusion for two Li$_{\rm x}$Ni$_{1/3}$Co$_{1/3}$Mn$_{1/3}$O$_{2}$ samples ($x=1$~and~0.3) using muon spin rotation/relaxation ($\mu^+$SR).

\begin{figure}[t]
  \begin{center}
    \includegraphics[keepaspectratio=true,width=125 mm]{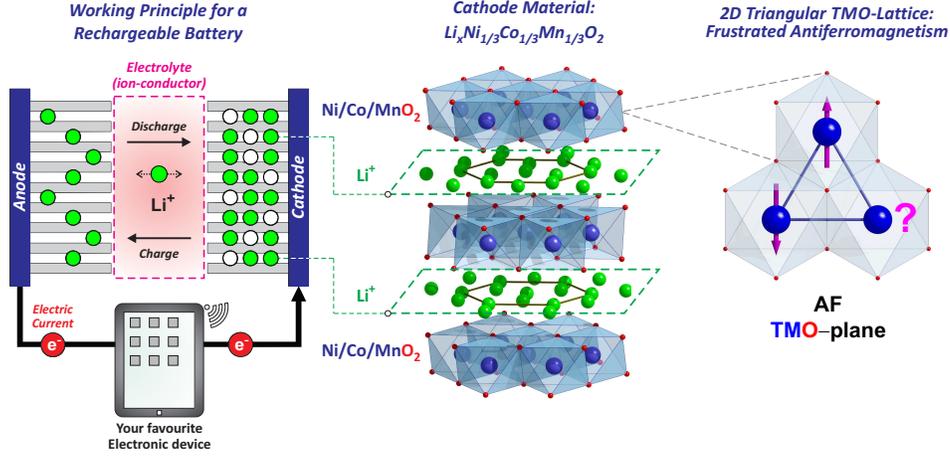}
  \end{center}
  \caption{Schematic figure showing the operational principle for a standard rechargeable battery (left). Middle panel show the archetypical atomic structure for a layered battery cathode materials. Finally, right panel show the triangular transition metal oxide (TMO) planes that creates the possibility for 2D frustrated antiferromagnetism (AF) within the planes. This figure highlights that the very same layered TMO compounds can be of interest for energy applications as well as low-temperature fundamental condensed matter physics.
    }
  \label{fig:structure}
\end{figure}

\section{Experimental Details}
Powder samples of Li$_{\rm x}$Ni$_{1/3}$Co$_{1/3}$Mn$_{1/3}$O$_{2}$ compound were prepared at Uppsala University, Sweden. For the low-temperature magnetic measurements, two samples ($x=1$~and~0.3) of approximately 1 g each were placed in small envelopes made of very thin Al-coated Mylar tape and then attached to a low-background, fork-type, sample holder. Subsequently, high-resolution $\mu^+$SR spectra were measured at the (continuous, DC) Swiss Muon Source (S$\mu$S), Paul Scherrer Institute, Villigen, Switzerland. By using the muon beamline $\pi$E1 with the Dolly spectrometer, zero-field (ZF) and weak transverse-field (wTF) spectra were collected for 2~K~$\leq{}T\leq$~150~K. For the ion-diffusion experiments, approximately 2 grams of the same samples were pressed into two discs with a 24~mm diameter and 1.5~mm thickness. Inside a helium glove-box the discs were packed into a Au-sealed (gold O-ring) powder cells made of pure titanium using a thin ($100~\mu$m) Kapton film as 'entrance window' for the muons. In addition, a silver mask with a hole matching the samples diameter was mounted onto the cell to ensure that the any minor background signal is non-relaxing over a wide temperature range. The cell was mounted onto the Cu end-plate of a liquid-He flow-type cryostat in the temperature range between 10 and 500~K. Subsequently, ZF-, wTF- and LF-$\mu^{+}$SR spectra were collected using the RIKEN-RAL / ARGUS spectrometer at the pulsed muon source ISIS/RAL in UK. The experimental techniques are described in more detail elsewhere \cite{muSR_book}.

\section{Results and Discussion: Low-temperature Magnetic Properties}
To obtain a first insight into the nature of the magnetic order in these compounds, we acquired zero-field (ZF) data at the lowest temperature ($T=2$~K). The $\mu^{+}$SR time spectrum show no indication of spontaneous muon precession, but rather only a fast exponentially decaying signal for both samples [see Fig.~2(a-b)]. Such signal indicate either a wide field-distribution, which is typical for spin-glasses (SG) or the presence of dynamical correlations. However, recent magnetometry measurements of our samples ($x=1$ and $x=0.3$) using zero-field-cooled (ZFC) and field-cooled (FC) protocols \cite{Wikberg_private} clearly show a divergence around $T=10$~K, which is a clear indication for a magnetically frustrated SG state. Further, information obtained also from related compounds using magnetometry as well as $\mu^+$SR \cite{Wikberg_PRB,Wikberg_JAP,Wikberg_muSR} clearly favours the static SG scenario . The ZF data was found to be well fitted to the sum of a very fast exponentially relaxing signal and a slower decaying stretched-exponential function:
\begin{eqnarray}
 A_0 \, P_{\rm ZF}(t) = A_{\rm fast}\cdot{}e^{-\lambda_{\rm fast} t~} + A_{\rm slow}\cdot{}e^{[(-\lambda_{\rm slow} t)^{n}]},
\label{eq:ZFfit}
\end{eqnarray}
This kind of stretched exponential relaxation is found to appear in a range of different systems, however, they are commonly connected to SG systems \cite{Phillips} containing geometrical frustration $e.g.$ the title compound. In similarity to our previous $\mu^+$SR studies of the closely related Li(Ni$_{0.8}$Co$_{0.1}$Mn$_{0.1}$)O$_2$ compound \cite{Wikberg_muSR}, we find that also the present compounds enter into a disordered spin-glass (SG) state at lowest temperature ($T=2$~K). Further, from the fits to Eq.~1 the critical exponent $n$ reaches a value $n\leq1/2$ [$n_{x=1}\approx0.41$ and $n_{x=0.3}\approx0.44$], indicating a decreased dimensionality (most likely 2D, considering the layered structure). From ZF measurements at $T=30$~K as well as $T=100$~K it is evident that both samples are in a paramagnetic state above at least $T=30$~K [see insets Fig.~2(a-b)].

To further investigate the PM to SG phase-transition we also performed detailed weak-transverse field (wTF~=50~G) measurements as a function of temperature. The wTF $\mu^{+}$SR time spectra for selected temperatures are shown in Fig.~2(c-d). Around $T=10$~K a total suppression of the externally applied field is seen as annihilation of the wTF asymmetry ($A_{\rm TF}$). The data is found to be well fitted by a combination of an exponentially relaxing cosine oscillation and a fast relaxing component (the latter becomes dominant at and below the transition):
\begin{eqnarray}
 A_0 \, P_{\rm TF}(t) = A_{\rm TF}\cos(2\pi f_{\rm TF}\cdot{}t+\phi_{\rm TF})\cdot{}e^{-\lambda_{\rm TF} t} + A_{\rm fast}\cdot{}e^{-\lambda_{\rm fast} t~}~,
\label{eq:ZFfit}
\end{eqnarray}
From such fits, the temperature dependence of $A_{\rm TF}$ can be extracted and a better picture of the transition is obtained [Fig.~2(e-f)]. For both samples it is clear that a magnetic phase transition into the previously mentioned SG state occurs at $T_{\rm SG}\approx12$~K where $A_{\rm TF}$ drops to zero very rapidly. This is in fact rather different from the Li(Ni$_{0.8}$Co$_{0.1}$Mn$_{0.1}$)O$_2$ compound \cite{Wikberg_muSR} where we found a gradual transition over a 50~K wide temperature range. Hence, contrary to the cluster glass like state found for the previous compound, Li$_{\rm x}$Ni$_{1/3}$Co$_{1/3}$Mn$_{1/3}$O$_{2}$ is believed to enter into a simple SG phase below $T_{\rm SG}$. This could possibly be explained by the equal distribution of transition metals creating a higher degree of disorder on the TMO planes. Worth noticing is also that the $x=1$ sample seems to have a slightly wider transition than $x=0.3$. This could be an artefact related to too low statistics for the wTF data. Another explanation could be that stoichiometric $x=1$ samples are well known to be notoriously difficult to synthesize.

\begin{figure}[t]
  \begin{center}
    \includegraphics[keepaspectratio=true,width=135 mm]{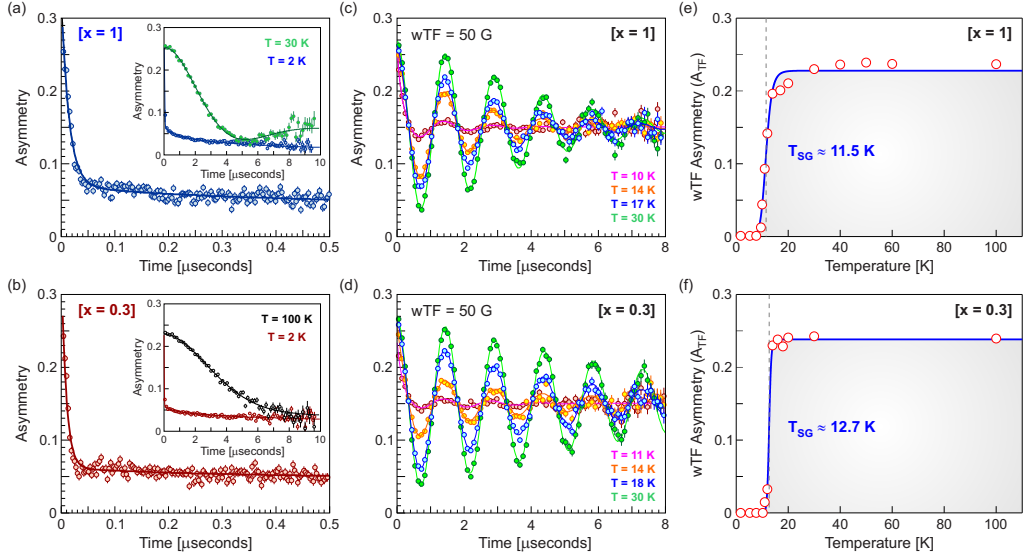}
  \end{center}
  \caption{\textbf{Low-temperature magnetic measurements at S$\mu$S continuous muon source.}
            (a-b) Zero-field (ZF) $\mu^+$SR spectra at $T=2$~K from the $x=1$ and $x=0.3$ samples, respectively. Data are shown in the very short time-scale in order to emphasize the fast relaxing signal. Solid line is a fit to Eq.~1. Insets show data in the longer time-scale together with a spectra at higher temperature, which in both cases display clear Kubo-Toyabe behaviour.
            (c-d) weak-transverse field (wTF~=~50~G) data from the $x=1$ and $x=0.3$ samples, respectively, for a series of temperatures across the magnetic transition. Solid lines are fits to Eq.~2.
            (e-f) Temperature dependence of the wTF asymmetry ($A_{\rm TF}$) from the two samples. A clear entrance into a magnetic state is observed in both cases for
            $T_{\rm SG}\approx12$~K.}
  \label{fig:Mag}
\end{figure}

\section{Results and Discussion: High-temperature Li-ion Diffusion}
In order to investigate Li-ion diffusion in these materials we have used our previously presented $\mu^+$SR technique for studying solid state ion diffusion in battery related materials \cite{Sugiyama_03,Mansson}. In similarity to our previous investigations of both Li-ion \cite{Sugiyama_03,Sugiyama_04,Sugiyama_05,Sugiyama_06,Sugiyama_07} and Na-ion \cite{Mansson} diffusion, a series of ZF, wTF~=~30~G as well as LF~=~5~G and 10~G $\mu^{+}$SR spectra were acquired in the temperature range between 50 and 500~K. At each temperature, the ZF and two LF spectra were fitted by an exponentially relaxing dynamic Kubo-Toyabe (KT) function plus a small background signal from the fraction of muons stopped mainly in the silver mask:
\begin{eqnarray}
 A_0\,P(t) &=&
  A_{\rm KT} G^{\rm DGKT}(\Delta, \nu, t)\exp(-\lambda t)+ A_{\rm BG}
\label{eq:DKT}
\end{eqnarray}
Furthermore, a global fitting procedure was employed over the entire temperature range using a common alpha ($\alpha$) and background asymmetry ($A_{\rm BG}$), but with temperature dependent field fluctuation rate ($\nu$) and relaxation rate ($\lambda$). The static width of the local field distribution ($\Delta$) was from individual fitting found to be more or less independent of temperature and was finally also treated as a common parameter for the global fit over the entire temperature range for each sample [$\Delta_{x=1}\approx0.35\cdot10^6$~s$^{-1}$ and $\Delta_{x=0.3}\approx0.24\cdot10^6$~s$^{-1}$].

\begin{figure}[t]
  \begin{center}
    \includegraphics[keepaspectratio=true,width=140 mm]{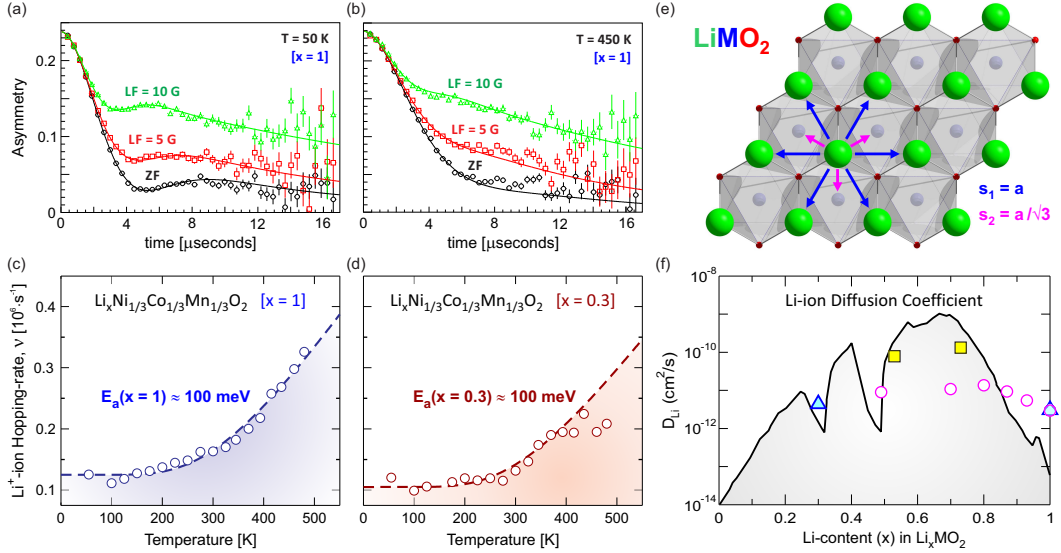}
  \end{center}
  \caption{\textbf{High-temperature diffusion measurements at ISIS pulsed muon source.}\\
            Zero-field (ZF) and Longitudinal-field (LF = 5 and 10 G) $\mu^+$SR time spectra collected at (a) $T=50$~K and (b) $T=450$~K for the $x=1$ sample. Solid lines are fits to Eq.~3.
            (c-d) Temperature dependence of the Li-ion hopping-rate [$\nu(T)$] obtained from our $\mu^+$SR experiments for $x=1$ and $x=0.3$ samples, respectively.
            (e) Li$^+$-ion jump-paths and distances within the lithium planes.
            (f) Resulting Li-ion diffusion coefficient ($D_{\rm Li}$) for the current two Li$_{\rm x}$Ni$_{1/3}$Co$_{1/3}$Mn$_{1/3}$O$_{2}$ samples shown as filled blue triangles. Also shown are results from previous $\mu^+$SR measurements on similar samples with different $x$ \cite{Sugiyama_08} (open circles) as well as from the Li$_x$CoO$_2$ compound \cite{Sugiyama_03} (filled yellow squares) together with a calculated curve for $D_{\rm Li}$ in Li$_x$CoO$_2$ as solid black line \cite{VanderVen}.
    }
  \label{fig:Diff}
\end{figure}

Below $T\approx120$~K, the spectra display a clear static behavior as shown for instance by the $T=50$~K spectra for the $x=1$ samples in Fig.~3(a). However, above $T_{\rm diff}\approx125$~K clear dynamic contribution sets in, which increases strongly with temperature, as shown in the $T=450$~K spectra for the $x=1$ sample [Fig.~3(b)]. The temperature dependence of the Li-ion hopping rate [$\nu(T)$] obtained from fitting both data sets to Eq.~3 are shown in Fig.~3(c-d) for the $x=1$ and $x=0.3$ samples, respectively. As seen, $\nu(T)$ displays a clear diffusive behavior above $T_{\rm diff}\approx150$~K for both samples. By fitting $\nu(T)$ to an Arrhenius type equation [dashed line in Fig.~3(c-d)], it is possible to extract the activation energy $E_{\rm a}\approx100$~meV for both samples. If we assume that $\nu(T)$ is a direct measure of the Li-ion hopping rate we can then express the diffusion coefficient $D_{\rm Li}$ as \cite{Borg}:
\begin{eqnarray}
D_{\rm Li}&=&
\sum^{n}_{i=1}\frac{1}{N_i}Z_{i}s_i^2\nu,
\label{eq:DLi}
\end{eqnarray}
Here, $N_i$ is the number of possible Li jump sites for the $i$:th path, $Z_i$ is the vacancy fraction, $s_i$ is the jump distance and $\nu$ is the hopping rate obtained from our $\mu^+$SR experiments. For the current samples there are in similarity to Li$_x$CoO$_2$ samples 2 different jumping paths to either a partly occupied Li-ion site ($N_1=6$, $Z_1=1-x$ and $s_1=$~\textbf{a}-axis~=~2.86347(2)~{\AA}) or to an empty interstitial site ($N_2=3$, $Z_2=1$ and $s_2=$~\textbf{a}-axis$/\sqrt{3}$) [see also Fig.~3(e)]. As a result we can calculate $D_{\rm Li}$ at 300 K for our two samples as $D_{\rm Li}^{300\rm K}[x=1]=3.5\cdot10^{-12}$~cm$^{2}$s$^{-1}$ and
$D_{\rm Li}^{300\rm K}[x=0.3]=5\cdot10^{-12}$~cm$^{2}$s$^{-1}$, respectively. If we insert such values in a graph showing $D_{\rm Li}$ from previous measurements of
Li$_{\rm x}$Ni$_{1/3}$Co$_{1/3}$Mn$_{1/3}$O$_{2}$ samples \cite{Sugiyama_08} we clearly see that the value for $x=1$ is fully reproducible. Our new data for $x=0.3$ falls where one can expect considering the previous measurements of $x=0.49$ samples. Please note that a different sample provider was used for the current study as compared to previously published data \cite{Sugiyama_08}.

An interesting detail appears when looking at the first principle calculations for $D_{\rm Li}$ in the closely related Li$_x$CoO$_2$ compound \cite{VanderVen} [see solid black line in Fig.~3(f)], several sharp minima at \emph{e.g.} $x=0.33$ and $x=0.5$ are visible. Such minima are related to different type of Li-vacancy ordering. However, from the present and previous measurements on Li$_{\rm x}$Ni$_{1/3}$Co$_{1/3}$Mn$_{1/3}$O$_{2}$ for a series of different Li-content, $x$, it seems clear that no such minima are present in this compound. This is probably reasonable since the disorder in the MTMO planes would create an unfavorable environment for the spontaneous formation of any Li-vacancy ordering.

\section{Summary}
By the use of a positive muon-spin rotation and relaxation ($\mu^{+}$SR) technique we have investigated the mixed transition metal oxide battery cathode material
Li$_{\rm x}$Ni$_{1/3}$Co$_{1/3}$Mn$_{1/3}$O$_{2}$. Low-temperature ($T=2-100$~K) measurements show that this compounds enters into a spin-glass like magnetic state below approximately $T=12$~K for both lithium contents $x=1$ and $x=0.3$, respectively. High-temperature ($T=50-500$~K) investigations show details regarding the Li-ion diffusion process in these compounds. Above $T_{\rm diff}\approx125$~K the lithium ions start to diffuse as seen from the exponential increase of the hopping rate $\nu(T)$. We are able to extract the activation energy for the diffusion mechanism as $E_{\rm a}\approx100$~meV, which seems to be approximately independent of the lithium content $x$.

\section{Acknowledgments}
This work was performed using the DOLLY spectrometer at the Swiss Muon Source (S$\mu$S) of the Paul Scherrer Institut (PSI), Villigen, Switzerland as well as the RIKEN-RAL / ARGUS spectrometer at the pulsed muon source ISIS/RAL in UK. We are very grateful for the support obtained during our experiments from the beamline and technical staff of the two facilities. This research was financially supported by the Swedish Research Council (VR), Swedish Energy Agency, Knut and Alice Wallenberg Foundation (KAW), the Swiss National Science Foundation (SNSF), MEXT KAKENHI Grant No. 23108003 and JSPS KAKENHI Grant No. 26286084.

\section*{References}

\begin{thebibliography}{21}

\bibitem{Mizushima}
K. Mizushima $et~al.$,
Mater. Res. Bull. \textbf{15}, 783 (1980).

\bibitem{Li_review}
Hong Li, Zhaoxiang Wang, Liquan Chen, and Xuejie Huang,
Adv. Mater. \textbf{21}, 4593 (2009).

\bibitem{VanderVen}
A. Van der Ven $et~al.$,
Phys. Rev. B \textbf{58}, 2975 (1998).

\bibitem{Seguin}
L. Seguin $et~al.$,
J. Power Sources \textbf{81/82}, 604 (1999).

\bibitem{Chebiam}
R.V. Chebiam $et~al.$
J. El.Chem. Soc. \textbf{148}, A49 (2001).

\bibitem{Yoshio}
M. Yoshio $et~al.$,
J. Pow. Sour. \textbf{90}, 176 (2000).

\bibitem{Choi}
J. Choi $et~al.$,
J. Pow. Sour. \textbf{162}, 667 (2006).

\bibitem{Ohzuku}
T. Ohzuku $et~al.$,
Chem. Lett. \textbf{7}, 642 (2001).

\bibitem{Dahbia}
M. Dahbia $et~al.$,
Electrochimica Acta \textbf{54}, 3211 (2009).

\bibitem{Anderson}
P. W. Anderson,
Mater. Res. Bull. \textbf{8}, 153 (1973).

\bibitem{Wikberg_PRB}
J. M. Wikberg $et~al.$,
Phys. Rev. B \textbf{81}, 224411 (2010).

\bibitem{Wikberg_JAP}
J. M. Wikberg $et~al.$,
J. Appl. Phys. \textbf{180}, 083909 (2010).

\bibitem{Wikberg_muSR}
J. M. Wikberg $et~al.$,
Phys. Proc. \textbf{30}, 202 (2012).

\bibitem{muSR_book}
Alain Yaouanc and Pierre Dalmas de Reotier,
\emph{Muon Spin Rotation, Relaxation, and Resonance}
(Oxford University Press, USA, 2011).

\bibitem{Wikberg_private}
J. M. Wikberg
\textit{Private Communication} (2014).

\bibitem{Phillips}
C. Phillips,
Rep. Prog. Phys. \textbf{59}, 1133 (1996).

\bibitem{Sugiyama_03}
J. Sugiyama $et~al.$,
Phys. Rev. Lett. \textbf{103}, 147601 (2009).

\bibitem{Mansson}
M. Mansson and J. Sugiyama,
Physica Scripta \textbf{88}, 068509 (2013).

\bibitem{Sugiyama_04}
J. Sugiyama $et~al.$,
Phys. Rev. B \textbf{82}, 224412 (2010).

\bibitem{Sugiyama_05}
J. Sugiyama $et~al.$,
Phys. Rev. B \textbf{84}, 054430 (2011).

\bibitem{Sugiyama_06}
J. Sugiyama $et~al.$,
Phys. Rev. B \textbf{85}, 054111 (2012).

\bibitem{Sugiyama_07}
J. Sugiyama $et~al.$,
Phys. Rev. B \textbf{87}, 024409 (2013).

\bibitem{Sugiyama_08}
J. Sugiyama $et~al.$,
Phys.Chem. Chem. Phys. \textbf{15}, 10402 (2013).

\bibitem{Borg}
R. J. Borg and G. J. Dienes,
\emph{An Introduction to Solid State Diffusion}
(San Diego, CA: Academic, 1988).

\end{thebibliography}

\end{document}